\documentclass[12pt]{article}
\usepackage{graphicx,floatflt,amssymb,epsfig,epsf,amsmath} 
\begin{document}
\renewcommand{\thesection}{\Roman{Section}}

\begin{center} 
{\Large \bf Many-body approach to low-lying collective excitations in a BEC approaching collapse}\\
\vspace*{1cm} 
\renewcommand{\thefootnote}{\fnsymbol{footnote}} 
{\large {\sf Anindya Biswas}\footnote{e-mail : anindya.biswas@ymail.com} and {\sf Tapan Kumar Das}\footnote{e-mail : kumartd@rediffmail.com} } \\ 
\vspace{10pt} 
{\small 
   {\em Department of Physics, University of Calcutta, 92 A.P.C. 
        Road, Kolkata 700009, India}}
 
\normalsize 
\end{center} 

\begin{abstract} 
An approximate many-body theory incorporating two-body correlations has been employed to calculate low-lying collective multipole frequencies in a Bose-Einstein condensate containing $A$ bosons, for different values of the interaction parameter $\lambda=\frac{Aa_{s}}{a_{ho}}$. Significant difference from the variational estimate of the Gross-Pitaevskii equation has been found near the collapse region. This is attributed to two-body correlations and finite range attraction of the realistic interatomic interaction. A large deviation from the hydrodynamic model is also seen for the second monopole breathing mode and the quadrupole mode for large positive $\lambda$.
\vskip 5pt \noindent 
\end{abstract} 

\hspace{1cm}
Low-lying collective excitations of a Bose-Einstein condensate can provide valuable information about the interactions and stability of the condensate, as the dimensionless interaction parameter $\lambda=\frac{Aa_{s}}{a_{ho}}$ decreases from a large positive (repulsive interaction) value to negative (attractive interaction) values. Starting from the Gross-Pitaevskii $(GP)$ equation, and using a sum rule approach, Stringari \cite{stringari} obtained analytic expressions in terms of expectation values of kinetic energy ($E_{kin}$) and harmonic confining potentials $(E_{ho})$, for the monopole $(\omega_{M})$ and quadrupole $(\omega_{Q})$ frequencies. In Fig.$1$ of Ref.$[1]$, variational estimates \cite{baym,fetter} were plotted against $\lambda$, which shows a sharp change for both $\omega_{M}$ and $\omega_{Q}$ as $\lambda$ approaches the variational critical value $\lambda_{cr}^{GP,var}=-0.671$ \cite{fetter}. Note that this provides an 
upper bound. Furthermore, numerical calculation of $E_{kin}$ by the $GP$ equation close to the collapse region involves large errors. Hence, it is desirable to calculate the low-lying excitation frequencies directly using a many-body approach, which can handle a finite range interaction instead of the contact interaction in the $GP$ equation.\\
\hspace*{1cm}
In the present work, we calculate low-lying collective excitation frequencies of a dilute Bose-Einstein condensate $(BEC)$, using the many-body potential harmonics expansion $(PHE)$ approach \cite{das}. Only two-body correlations are relevant in the dilute many-body system. Hence the $(ij)$-Faddeev component of the many-body wave function, is a function of the relative separation $(\vec{r}_{ij})$ and a global length called hyperradius $(r)$ only. In the $PHE$ method, one expands the Faddeev component $(\psi_{ij})$, corresponding to the $(ij)$-interacting pair of the condensate containing $A$ atoms, in the corresponding potential harmonics $(PH)$ basis \cite{fabre} (which is the subset of the full hyperspherical harmonics $(HH)$ basis \cite{ballot}, needed for the expansion of the two-body potential $V(\vec{r}_{ij})$)
\begin{equation}
 \psi_{ij}(\vec{r}_{ij},r)=r^\frac{-(3A-4)}{2}\Sigma_{K}\textit{P}_{2K+l}^{lm}(\Omega_{ij})u_{K}^{l}(r).
\end{equation}
Here, $l$ and $m$ are the orbital angular momentum of the system and its projection and $\Omega_{ij}$ represents the full set of hyperangles for the $(ij)$-partition. A closed analytic expression can be obtained for the $PH$, $\textit{P}_{2K+l}^{lm}(\Omega_{ij})$ \cite{fabre}. The expansion $(1)$ is, in general, very slow due to the fact that the lowest order $PH$ is a constant and does not represent the strong short-range correlation of the interacting pair, arising from the very short range repulsion of the interatomic interaction. To enhance the expansion we introduce a short-range correlation function $\eta(r_{ij})$, in analogy with atomic systems \cite{lin,lin2}. This is obtained as the zero energy solution of the two-body Schr\"{o}dinger equation with the chosen two-body potential \cite{barletta}, corresponding to the appropriate s-wave scattering length $(a_{s})$. Thus we replace expansion $(1)$ by \cite{das1} 
\begin{equation}
 \psi_{ij}(\vec{r}_{ij},r)=r^\frac{-(3A-4)}{2}\Sigma_{K}\textit{P}_{2K+l}^{lm}(\Omega_{ij})u_{K}^{l}(r)\eta(r_{ij}).
\end{equation}
Substitution of this expansion in the many-body Schr\"{o}dinger equation and projection on a particular $PH$ gives rise to a system of coupled differential equation $(CDE)$ in $r$ \cite{das,das1}, which is solved numerically using hyperspherical adiabatic approximation $(HAA)$ \cite{das2}. Details of the procedure can be found in ref. \cite{das1}.\\
\hspace*{1cm}
In the $HAA$, the coupling potential matrix together with the diagonal hypercentrifugal repulsion is diagonalised to get the effective potential, $\omega_{o}(r)$, as the lowest eigenvalue of the matrix for a particular value of $r$. Collective motion of the condensate in the hyperradial space takes place in the effective potential $\omega_{o}(r)$. Ground state in this well gives the ground state energy $(E_{00})$ of the condensate corresponding to $n=0$, $l=0$. Here $E_{nl}$ is the energy in oscillator units $(o.u.)$ of the $n^{th}$ radial excitation of the $l^{th}$ surface mode. Hyperradial excitations corresponding to the breathing mode for $l=0$, give the monopole frequencies as $\omega_{Mn}=(E_{n0}-E_{00})$. For $l\neq0$, we get the surface modes. Numerical calculation of the off-diagonal potential matrix elements for $l\neq0$ is fraught with large inaccuracies and its numerical computation is very slow. On the other hand the diagonal hypercentrifugal term is very large for large $A$ and contributes most to the potential matrix. Hence we disregard $l>0$ contributions to the off-diagonal matrix elements. For non-zero orbital angular momentum, we thus get the effective potential $\omega_{l}(r)$ in the hyperradial space. The ground state in this potential is the ground state of the $l^{th}$ surface mode $E_{0l}$ and $n^{th}$ radial excitations provide $E_{nl}$. The lowest monopole and quadrupole frequencies are given by $\omega_{M}=E_{10}-E_{00}$ and $\omega_{Q}=E_{02}-E_{00}$. The $s$-wave scattering length ($a_{s}$) has been chosen as $2.09783\times10^{-4}$ $o.u.$ for repulsive interaction and $-1.39217\times10^{-4}$ $o.u.$ for attractive interaction. The interatomic interaction is chosen as the van der Waals potential with a hard core of radius $r_{c}$, $viz.$, $V(r_{ij})=\infty$ for $r_{ij}<r_{c}$ and $=-\frac{C_{6}}{r_{ij}^{6}}$ for $r_{ij} \geq r_{c}$. The value of $C_{6}$ is chosen to be appropriate for rubidium atoms. The value of $r_{c}$ is adjusted to get the desired value of $a_{s}$ \cite{pethick}. All quantities are expressed in oscillator units ($o.u.$) appropriate for the JILA experiment with $^{85}Rb$ \cite{roberts}. The $PHE$ equation is then solved numerically for different $A$ to calculate $E_{nl}$ for different values of $\lambda$.\\
\hspace*{1cm}
In Fig.$1$ we plot calculated $\omega_{M}$ and $\omega_{Q}$ as a function of the dimensionless interaction parameter $\lambda$. For comparison we include the variational estimates from the $GP$ equation \cite{stringari}. We have also included results of a numerical solution of the $GP$ equation \cite{tiwari}. However calculation of $E_{kin}$ and $E_{int}$ using the $GP$ program have large errors for values of $\lambda$ close to collapse for attractive condensates as also for large positive $\lambda$. This is shown by the lack of agreement with the virial identity involving $E_{kin}$, $E_{int}$ and $E_{ho}$, $viz$. $2E_{kin}-2E_{ho}+3E_{int}=0$ \cite{dalfovo}. Hence in Fig.$1$, the curves obtained by a numerical solution of the $GP$ equation is shown only in the region where it is reliable. It is seen that for positive values of $\lambda$, all the three curves for the monopole frequency are close to each other. Results of $PHE$ and variational estimates are almost indistinguishable. However for negative $\lambda$, $\omega_{M}$ by the $PHE$ method drops faster as $\lambda$ approaches the critical value $\lambda_{cr}=-0.461$. Note that while $GP$ equation predicts a critical value of $-0.571$ (variational estimate is $-0.671$) \cite{dalfovo}, experimental critical number is $-0.459\pm0.012\pm0.054$ \cite{roberts}. The prediction by the $PHE$ method agrees well with this value \cite{kundu}. As the attractive condensate approaches collapse, with $\lambda$ approaching $\lambda_{cr}$ from the right, the condensate becomes softer (more compressible) and the breathing mode easier.\\
\hspace*{.5cm}
\begin{figure}[hbpt]
\vspace{-10pt}
\centerline{
\hspace{-3.3mm}
\rotatebox{0}{\epsfxsize=12cm\epsfbox{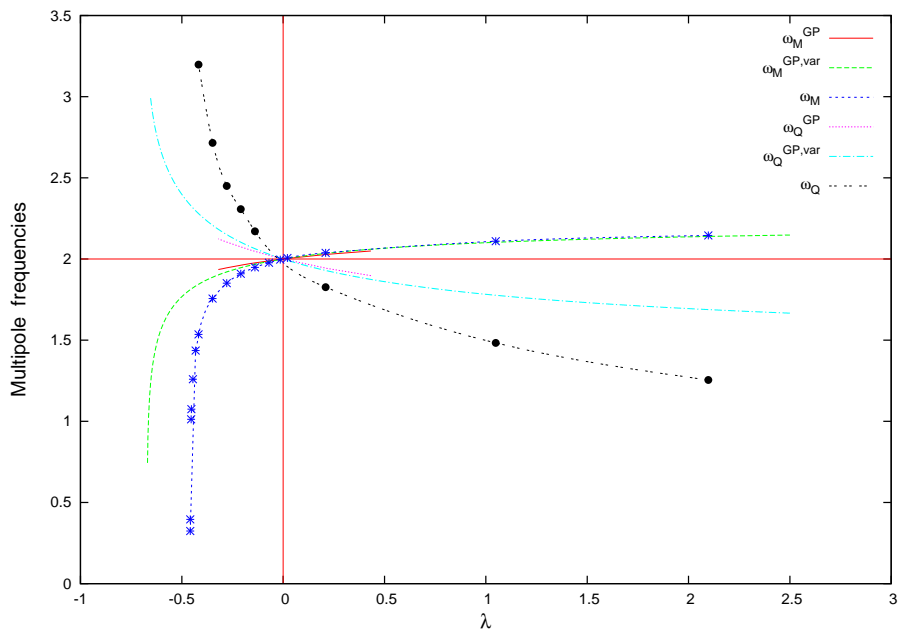}}}
\caption{Calculated monopole $(\omega_{M})$ and quadrupole $(\omega_{Q})$ frequencies as functions of 
$\lambda$. Superscripts GP, GP,var and PHE correspond respectively to numerical solution of the GP equation, 
its variational estimate, and the present calculation.}
\end{figure}

\hspace*{1cm}
On the other hand the quadrupole frequency $(\omega_{Q})$ shows a greater deviation from the variational 
estimate -- both for positive and negative values of $\lambda$. Since $\lambda_{cr}=-0.461$, it is natural 
that both monopole and quadrupole frequencies calculated by $PHE$ method have rapid changes as $\lambda$ 
approaches $-0.461$. The difference between the collective excitation frequencies calculated by $PHE$ and 
$GP$ equation can be attributed to the two-body correlations and finite range interactions. As the 
condensate approaches collapse, its size decreases rapidly \cite{kundu}, so that the finite range 
interatomic attractions of the realistic atom-atom potential dominate and correlations play a more important 
role. Although for small values of $|\lambda|$ use of an effective interaction in terms of $a_{s}$ is 
adequate, the strong correlations arising from the attractive part of the realistic atom-atom interaction 
produces discernible effects \cite{blume,gao,geltman,kalas} for $\lambda$ approaching $\lambda_{cr}$. One 
can also qualitatively interpret the dependence of $\omega_{Q}$ on $\lambda$ as follows. The two-body 
correlation disallows proximity of two interacting atoms in the repulsive case. Again in the $l=2$ state the 
atoms are further apart compared to the $l=0$ state. It is therefore easier ($i.e.$ excitation energy is 
lower) for the condensate to jump from $l=0$ to $l=2$ state, when two-body correlations are included than 
when they are excluded (as in the $GP$ equation). The inverse is true for the attractive case. The results 
obtained in the present work are in conformity with these arguments. Further discussions on the behaviour of 
$\omega_{Q}$ for large positive $\lambda$ will be taken up together with those of higher excitation 
frequencies, after we present them.\\
\hspace*{.5cm}
\begin{figure}[hbpt]
\vspace{-10pt}
\centerline{
\hspace{-3.3mm}
\rotatebox{0}{\epsfxsize=12cm\epsfbox{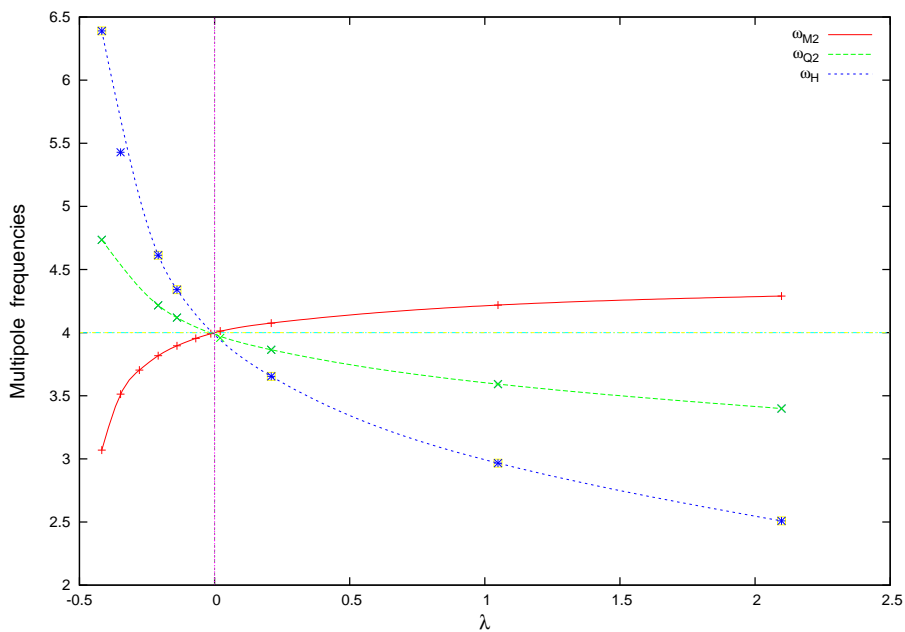}}}
\caption{Higher order multipole frequencies calculated by PHE method, as functions of $\lambda$.}
\end{figure}

\hspace*{1cm}
In Fig.$2$ we present the multipole frequencies which are degenerate with a value of $4$ in the non-interacting case, $viz.$, $\omega_{M2}=E_{2,0}-E_{0,0}$, $\omega_{Q2}=E_{1,2}-E_{0,0}$ and $\omega_{H}=E_{0,4}-E_{0,0}$. As expected from the softness of the condensate , the second breathing mode frequency $\omega_{M2}$ is larger than $4$ for positive $\lambda$ and less than $4$ for negative $\lambda$, decreasing sharply as $\lambda \rightarrow \lambda_{cr}$. This once again shows that the condensate becomes softer as $\lambda \rightarrow \lambda_{cr}$, and harder and more rigid as $\lambda$ increases to large positive values. On the ther hand, both $\omega_{Q2}$ and $\omega_{H}$ are less than $4$ for $\lambda > 0$ and greater than $4$ for $\lambda < 0$. Both increase sharply as $\lambda \rightarrow \lambda_{cr}$, $\omega_{H}$ increasing much faster than $\omega_{Q2}$. As positive $\lambda$ increases, the condensate becomes less compressible (more rigid) costing less energy for rotational excitations. Hence the spacing between the rotational levels decrease. For negative $\lambda$ close to $\lambda_{cr}$, $\omega_{H}$ is much larger than $\omega_{Q2}$. This shows that for such soft condensates, the first breathing mode of $l=2$ surface mode is easier to excite than the ground state of the $l=4$ surface mode, although they are degenerate in the non-interacting limit. This is consistent with the general picture and intuition. One can also note that $\omega_{Q2}-\omega_{Q}=E_{1,2}-E_{0,2}$ decreases rapidly from $2$ as $\lambda$ decreases from zero to $\lambda_{cr}$. This again shows the softness of the $l=2$ surface mode of the attractive condensate.\\
\hspace*{1cm}
We next come back to a discussion of our results as compared with those by the $GP$ equation and the hydrodynamic model, as also with available experimental results. As mentioned earlier, results using the numerical solution \cite{tiwari} of the $GP$ equation are not reliable for almost all values of $\lambda$ except small $|\lambda|$. Moreover $\omega_{M}$ and $\omega_{Q}$ given in terms of $E_{kin}$ and $E_{ho}$ are rigorous upper bounds for the frequency of the lowest states excited by the appropriate multipole operator \cite{stringari}. The upper bounds get further upwardly shifted for the variational estimate. The $PHE$ result for $\omega_{Q}$ is lower than the variational estimate; hence it is consistent with this requirement. Using the hydrodynamic model Stringari also derived a dispersion law for the frequencies of the normal modes \cite{stringari}, $viz.$, $\omega(n,l)=\omega_{0}(2n^{2}+2nl+3n+l)^{1/2}$, where $\omega_{0}$ is the frequency of the spherically symmetric trap. This is valid in the limit $\lambda>>1$, where kinetic energy contribution can be 
disregarded. While the $PHE$ result for $\omega_{M}$ is very close to this estimate, that for $\omega_{Q}$ is close to but less than this limit. Available experimental results \cite{jin,mews,kurn,onofrio} are for axially symmetric traps only -- measured $\omega$ for $m=\pm2$ and $m=0$ agree well with the $GP$ equation and the hydrodynamic model estimates for the deformed trap. Since accuracy of the hydrodynamic model 
estimate is expected to decrease as $n$ and $l$ increase, due to the neglect of the kinetic energy 
contribution \cite{stringari} and $\omega_{M}$ calculated by $PHE$ agrees remarkably well with that from the 
$GP$ equation, it is important to have accurate measurement of $\omega_{Q}$ for the spherical trap. The 
experimental procedure of ref.~\cite{onofrio} can be utilized for this purpose, since it is suitable for any 
geometry of the trap. This technique uses the optical dipole force of a rapidly scanning laser beam to 
excite surface modes by inducing deformations of the trap potential.\\
\hspace*{1cm}
More remarkable is the disagreement of the second breathing mode ($\omega_{M2}$). The $PHE$ predicts a value 
$>4$ for positive $\lambda$, while the $\lambda>>1$ limit by the hydrodynamic model gives $\sqrt{14}$. One 
expects that the system becomes less and less compressible as positive $\lambda$ increases. Consequently one 
would expect $\omega_{M2}>4$ for positive $\lambda$, which disagrees with the hydrodynamic model prediction. 
This is an indication that the kinetic energy contribution can no longer be disregarded for such excitation 
energies. To resolve the issue, one needs a precise measurement of $\omega_{M2}$. The higher mode can be 
excited by an appropriate frequency of the laser beam.\\
\hspace*{1cm}
In summary, we have investigated monopole, quadrupole and hexadecapole frequencies (corresponding to $l=0$, $2$ and $4$ respectively) of the collective motion of Bose-Einstein condensates by a many-body approach based on hyperspherical harmonics method. We include the most important two-body correlations, disregarding higher-body correlations. We notice that two-body correlations and the finite range attraction of a realistic two-body interaction make the attractive condensate softer compared with the use of a contact interaction. This is consistent with intuitive expectations. Consequently 
our results differ appreciably from those by the GP equation in the collapse region for an attractive 
condensate. The effect of two-body correlations become increasingly more relevant as the 
condensate approaches collapse. For repulsive condensates the monopole frequency agrees very well with the 
the GP estimate, while the quadrupole frequency differs by about 10\% of the asymptotic value. For the 
second breathing mode the PHE result is greater than the non-interacting value, as expected intuitively, 
for $\lambda>0$. This is in sharp contrast with the asymptotic ($\lambda >> 1$) GP estimate of $\sqrt{14}$, 
indicating that the disregard of kinetic energy terms in the hydrodynamical model is not valid for such  excitation levels. This calls for precise measurements of these multipole frequencies.\\
\hspace*{1cm}
This work has been partially supported by the Department of Science and Technology (DST, India) and the University Grants Commission (UGC India). AB acknowledges Payodhinath Mukherjee Research Scholarship of the University of Calcutta. TKD wishes to thank Prof. Sandro Stringari for suggesting the problem and very helpful discussions. He also wishes to thank the Center for $BEC$ studies at the University of Trento, Italy, for the warm hospitality during his visit.\\

\end{document}